# Time Isotropy, Lorentz Transformation and Inertial Frames

Somajit Dey

*Department of Physics, University of Calcutta, 92, A.P.C Road, Kolkata-700009, India*

Email: sdphys_rs@caluniv.ac.in and dey.somajit@gmail.com

Homogeneity of space and time, spatial isotropy, principle of relativity and the existence of a finite speed limit (or its variants) are commonly believed to be the only axioms required for developing the special theory of relativity (Lorentz transformations). In this paper it is shown, however, that Lorentz transformation cannot actually be derived without the explicit assumption of time isotropy (time-reversal symmetry) which is logically independent of the other postulates of relativity. Postulating time isotropy also restores the symmetry between space and time in the postulates of relativity. Inertial frames are defined in influential texts as frames having space-time homogeneity and spatial isotropy only. Inclusion of time isotropy in that definition is also suggested.

## I. INTRODUCTION

Since its advent, the foundations of special theory of relativity (STR) have often been reviewed in order to find the minimum number of axioms required to develop the theory rigorously [1]. These minimalist attempts claim to have founded STR on the assumptions of *homogeneity of space and time* (STH), *spatial isotropy* (SI) and *principle of relativity* (PR), thus making Einstein's 2$^{nd}$ postulate [2], that of the constancy of speed of light in free space, redundant. Actually, these assumptions give two possible kinematics viz. the Galilean transformation (GT) and the Lorentz transformation (LT) [3]. From LT we know that a space coordinate and time mix symmetrically. With hindsight, therefore, it seems intriguing that time isotropy (TI) should be absent as a postulate of STR when SI is explicitly postulated and homogeneity is assumed in both space and time. By TI we mean the equivalence of the two time directions viz. past and future. TI is the symmetry by which all fundamental classical laws are time reversal invariant. Exclusion of TI becomes all the more puzzling when SI and STH are taken to be the only defining properties of inertial frames [4]. Inertial frames are devised to make the description of physics simplest. Space and time in these frames, therefore, should possess the highest possible symmetry. Since time is independent of space by definition, TI is independent of SI and STH. Hence, TI should also be a defining attribute of inertial frames along with SI and STH. Moreover, all the fundamental classical laws are time symmetric. This becomes natural if TI is considered a defining property of inertial frames (just as the rotational invariance of classical laws is considered a natural consequence of SI).

In the following we derive LT and GT from PR, STH, SI and TI. The purpose of this article is to point out that PR, STH, SI and the existence of a finite speed limit indeed do not exhaust all the postulates of STR, since LT for a pure boost cannot be logically derived from them alone without appealing to TI. How then did the authors of the abovementioned literature get to LT (or GT for that matter) without assuming TI? The logical fallacies that allowed them to do so will contextually be remarked upon in due course by comparing with our proposed development.

## II. DERIVING LT

### a. Definitions and axioms

By "frame" in the following, we mean inertial frames with their own Cartesian triads and time.

Inertial frames are defined as frames having STH, SI and TI.

PR is postulated to hold between inertial frames.

### b. 1$^{st}$ step: Linearity of transformations from STH

Physics deals with laws of nature which deal with change of state in turn. Description of change of state in any frame is in terms of space and time. Consider now any two such frames $S\,(x,y,z,t)$ and $S'(x',y',z',t')$ linked by single-valued transformation functions,

$$x' = x'(x,y,z,t), \qquad (1)$$
$$y' = y'(x,y,z,t), \qquad (2)$$
$$z' = z'(x,y,z,t) \text{ and} \qquad (3)$$
$$t' = t'(x,y,z,t), \qquad (4)$$

and their inverses. The three rectilinear space-coordinates and time are represented by their usual symbols. These functions must be differentiable everywhere in the space-time continuum (i.e. the domain of the transformation functions) because the existence of singular point(s) would violate STH (unless singular points span the entire domain). Differentiating Eq. (1)–(4),

$$\Delta x' = \frac{\partial x'}{\partial x}\Delta x + \frac{\partial x'}{\partial y}\Delta y + \frac{\partial x'}{\partial z}\Delta z + \frac{\partial x'}{\partial t}\Delta t, \qquad (5)$$

$$\Delta y' = \frac{\partial y'}{\partial x}\Delta x + \frac{\partial y'}{\partial y}\Delta y + \frac{\partial y'}{\partial z}\Delta z + \frac{\partial y'}{\partial t}\Delta t, \qquad (6)$$

$$\Delta z' = \frac{\partial z'}{\partial x}\Delta x + \frac{\partial z'}{\partial y}\Delta y + \frac{\partial z'}{\partial z}\Delta z + \frac{\partial z'}{\partial t}\Delta t \text{ and} \qquad (7)$$

$$\Delta t' = \frac{\partial t'}{\partial x}\Delta x + \frac{\partial t'}{\partial y}\Delta y + \frac{\partial t'}{\partial z}\Delta z + \frac{\partial t'}{\partial t}\Delta t. \qquad (8)$$

According to STH, no point in space or time is preferred. So no choice of origin is physically favoured and space and time intervals are the only concepts that have any physical relevance. Hence, equal space-time intervals in $S$ should correspond to equal space-time intervals in $S'$. This means that Eq. (5)-(8) must be independent of $(x,y,z,t)$ which is possible only if the transformation functions in Eq. (1)–(4) are linear in $(x,y,z,t)$. Rewriting Eq. (1)–(4) in matrix form thus we get

$$\begin{pmatrix} x' \\ y' \\ z' \\ t' \end{pmatrix} = \boldsymbol{T} \begin{pmatrix} x \\ y \\ z \\ t \end{pmatrix} + \begin{pmatrix} o_x \\ o_y \\ o_z \\ o_t \end{pmatrix}, \qquad (9)$$

where $\boldsymbol{T}$ is a transformation matrix independent of $(x,y,z,t)$, and the right most column matrix is a constant dependent on the choice of origin. For points fixed in $S'$ space, $\frac{dx'}{dt}=0, \frac{dy'}{dt}=0, \frac{dz'}{dt}=0$ as seen from $S$. Hence, differentiating the first three rows of Eq. (9) with respect to $t$,

$$\begin{pmatrix} 0 \\ 0 \\ 0 \end{pmatrix} = \boldsymbol{\Gamma} \begin{pmatrix} \frac{dx}{dt} \\ \frac{dy}{dt} \\ \frac{dz}{dt} \end{pmatrix} + \begin{pmatrix} n_x \\ n_y \\ n_z \end{pmatrix} \qquad (10)$$

where $\boldsymbol{\Gamma}$ is some matrix and the rightmost column is constant. Solving Eq. (10) for $\left(\frac{dx}{dt}\; \frac{dy}{dt}\; \frac{dz}{dt}\right)^T$ we find that $S'$ moves with a constant velocity with respect to $S$. Therefore, the inertial frames move with uniform velocity relative to each other.

### c. 2$^{nd}$ step: Setting up parallel triads

If SI and STH hold in every frame, we can choose any origin and orientation without loss of generality. First we choose coincident origins so that the rightmost column in Eq. (9) becomes null. Then, the negative $X'$ axis of the $S'$ frame triad is chosen along the relative velocity of $S$, as seen from $S'$. As seen from $S$, the space points fixed on $X'$ axis, in $S'$ space, move forever along the relative velocity of $S'$. From SI, $S$ also sees these points to be collinear and aligned along their direction of motion. Let us hold that the $X$ and $X'$ axes become parallel if we choose this direction (i.e. along the relative velocity of $S'$) to be the positive $X$ axis of $S$.

For different values of the positive constant $a$, let us choose $y = a$ lines parallel to $X$ axis that are ***fixed*** in $S$. All the points on these lines, as seen from $S'$, are ***moving*** while maintaining constant distances from $X'$ axis. These lines, therefore, lie on cylinders of different radii with the $X'$ axis as their central axis. Free particles moving uniformly in $X - Y$ plane and transverse to $y = a$ lines, should move uniformly, i.e. in straight lines, with respect to $S'$ too and intersect the $y = a$ lines as seen from $S'$. Hence, the $y = a$ lines in $S$ are seen to be coplanar in $S'$ as well. $S'$ now chooses the positive $Y'$ axis as orthogonal to $X'$ and lying in this plane and towards the $y = a$ lines. Let us hold that $Y$ and $Y'$ axes are now parallel. $S$ ($S'$) now, with its $X$ ($X'$) and $Y$ ($Y'$) axes set up, uses the right hand rule to find the $Z$ ($Z'$) axis. We have followed Ref. [5] for this set up of triads. As argued in Ref. [5], $S$ and $S'$ now are identical vector spaces, and can agree to rotate identically to some other chosen orientation.

The whole point of this exercise has been to set up *parallel* vector spaces. However, the high point to note is that $S$ and $S'$ were not set up equivalently. The bold faced italics above highlight these differences in defining the axes of $S$ and $S'$. The $Y$ axis in $S$ was chosen freely by referring to ***fixed*** collinear particles in $S$, while the $Y'$ axis in $S'$ was chosen by referring to particles ***moving*** with respect to $S'$. This implies that the frame, say $S''$, that $S$ transforms into while it is brought to rest relative to $S'$ without any other operation being performed, is not necessarily identical or parallel to $S'$. This operation on $S$ that transforms it into $S''$ is called a "boost". In other words, the transformation $S \to S''$ exists only by virtue of motion along the $X$ axis. But nothing in the set up of parallel vector space $S'$ implies that $S''$ and $S'$ must be parallel. For example, $S''$ might just be rotated with respect to $S'$ about the $X'$ axis. Noting this possible difference is important since the transformation $S \to S''$, and not the transformation $S \to S'$, constitutes an LT for a pure boost. To the best of our knowledge however, the possible differences between $S'$ and $S''$ have so far been overlooked in the literature. The unsuspecting authors therefore, have treated the two transformations $S \to S''$ and $S \to S'$ to be completely identical. This mistake adds a forbidden artefact in what should be a maximally general development towards deriving LT. As we will see, this is the reason why it appeared that TI is not required when developing STR. TI is actually *the* postulate which necessitates that $S'$ and $S''$ be parallel and hence, if they adopt the same units, become identical.

**d. 3$^{rd}$ step: Form of a pure boost**

Let the velocity of $S''$ (and hence $S'$) relative to $S$ be $\boldsymbol{v}$ along $X$ axis. From Eq. (9) as applied to $S \to S''$,

$$\begin{pmatrix} x'' \\ y'' \\ z'' \end{pmatrix} = \boldsymbol{\Gamma} \begin{pmatrix} x \\ y \\ z \end{pmatrix} + t \begin{pmatrix} c_x \\ c_y \\ c_z \end{pmatrix}. \quad (11)$$

$\boldsymbol{\Gamma}$ was introduced in Eq. (10). Since the left hand side of Eq. (11) is a vector (position vector in $S''$) so must be the right hand side. $\boldsymbol{\Gamma}$ therefore acts as a linear operator (this idea is inspired from Ref. [5]). The last column vector in Eq. (11) is independent of $(x, y, z, t)$. It, therefore, must be along the velocity vector (i.e. $\boldsymbol{v}$) of $S''$ relative to $S$ since all the other directions are equivalent by SI in $S$. The next problem is to find how the time of $S$ transforms into that of $S''$. From Eq. (9) again

$$t'' = et + \boldsymbol{V} \cdot \boldsymbol{r}, \quad (12)$$

where $e$ is a scalar and $\boldsymbol{V}$ is some vector independent of $\boldsymbol{r}$. By SI again, $\boldsymbol{V}$ must be along $\boldsymbol{v}$, i.e.

$$\boldsymbol{V} = f\boldsymbol{v} \quad (13)$$

for some scalar $f$ independent of $(x, y, z, t)$.

From SI, $\boldsymbol{v}$ must be an eigenvector of $\boldsymbol{\Gamma}$. The most general form of the linear operator $\boldsymbol{\Gamma}$ can therefore, be written as [5]

$$\boldsymbol{\Gamma} = \boldsymbol{v}b\boldsymbol{v}\cdot + a(1 - \boldsymbol{v}\boldsymbol{v}\cdot) + d\boldsymbol{v}\times \quad (14)$$

where $a$, $b$ and $d$ are scalars independent of $(x, y, z, t)$. The first term stretches the component of any vector along $\boldsymbol{v}$ and the second the component normal to $\boldsymbol{v}$. The third term rotates the vector about $\boldsymbol{v}$. Eq. (11) in vector form thus looks like

$$\boldsymbol{r}'' = a\boldsymbol{r} + (b - a)(\boldsymbol{v}\cdot\boldsymbol{r})\boldsymbol{v} + d\boldsymbol{v}\times\boldsymbol{r} + c\boldsymbol{v}t \quad (15)$$

It is interesting to note that this can also be thought of as $\boldsymbol{r}''$ being expressed in terms of 3 basis vectors $\boldsymbol{r}$, $\boldsymbol{v}$ and $\boldsymbol{v}\times\boldsymbol{r}$ in the 3-D space of $S$. Since $a$, $b$, $c$, $d$, $e$ and $f$ are all scalars, SI implies that they must be functions of the magnitude of relative velocity (i.e. $v$) and not its direction. For $\boldsymbol{r}''$ fixed in $S''$, $\frac{d\boldsymbol{r}}{dt} = \boldsymbol{v}$. Differentiating Eq. (15) with respect to $t$ then gives

$$a + (b - a)v^2 + c = 0. \quad (16)$$

Using this, written in coordinate form, Eq. (15) and (12) go as

$$x'' = -cx + cvt, \quad (17)$$
$$y'' = ay - dvz, \quad (18)$$
$$z'' = az + dvy, \quad (19)$$
$$t'' = et + fvx. \quad (20)$$

The transformation equations almost always appear in literature in these forms and rarely in their vector forms Eq. (15) and (12). In almost all the cases, however, $d$ is claimed to be zero by assuming that $Y$ axis of the boosted $S$ (i.e. $S''$) is parallel to $Y$ axis. This means that $Y''$ and $Y'$ are parallel since $Y$ and $Y'$ were originally set to be parallel. As should be clear by now, this assumption is wrong since $S'$ and $S''$ are not necessarily parallel (see last paragraph of Sec. (II.c)). Since this is a key point, let us reemphasise that *$Y'$ axis in $S'$ is deliberately chosen to be parallel to $Y$ in (unboosted) $S$.* ***But this in no way warrants that the $Y'$ axis of $S'$ will be parallel (or equivalent) to the $Y$ axis of the boosted $S$, viz. $S''$.*** Before the boost, $Y$ axis was ***moving*** relative to $S'$ and was parallel to $Y'$ by definition. Eq. (18-19) then tells us what $Y$ would look like (compared to $Y'$ which is parallel to $Y$) when it is ***brought to rest*** relative to $S'$. It won't look like $Y'$ if $d \neq 0$. Unless we can prove $d = 0$ from the axioms of STR, $d \neq 0$ should remain a perfectly valid possibility. No experiment however precise, can ever tell $d$ is exactly 0, $d$ might still be very small but 0. Should $d = 0$, it must be established from principles. The root of the prevalent conviction that this ***boosted*** $Y$ (viz. $Y''$) must remain parallel to $Y'$ (and hence $d$ must be equal to 0) lies in the logical mistake discussed in the last paragraph of Sec. (II.c). The previous authors took $S \to S'$ to give the LT for a pure boost, while they should have taken $S \to S''$ instead.

Another argument for showing $d = 0$ assumes that the transformation equations (18-19) remain the same if only $X$ and $X''$ are inverted keeping the other axes unchanged [6]. However, this is wrong too. Inverting $X$ implies that we set the $Z$ axis from $X$ and $Y$ by left-hand rule instead of the right-hand rule as discussed in Sec. (II.c). This also changes the sense of $\boldsymbol{v}\times\boldsymbol{r}$ in Eq. (15). Hence, Eq. (18-19) do not remain the same but $d$ there is effectively replaced by $-d$ if $X$ and $X''$ are inverted.

**That $d$ is indeed 0, can be established from TI.** A non-zero $d$ would violate TI because then, time reversal in $S$, i.e.

$$\boldsymbol{r} \to \boldsymbol{r}, t \to -t, \boldsymbol{v} \to -\boldsymbol{v} \quad (21)$$

would imply $t'' \to -t''$ alright but not $\boldsymbol{r}'' \to \boldsymbol{r}''$. So TI would be violated. In other words, the general transformation equation(s) for a boost would not remain invariant under time reversal in both the frames even though they would retain form under any spatial rotation (SI) or origin shift (STH). Hence, TI emerges as key to our development apart from the well-recognised axioms of SI and STH.

Note that the existence of a non-zero $d$ implies that a certain handedness is preferred to the other; right handedness being preferred for $d > 0$ and left handedness for $d < 0$. This asymmetry in handedness cannot be argued away by postulating continuous symmetries like SI or STH. Rather, a discrete symmetry like time isotropy (time reversal symmetry) is required.

The requirement of time reversal symmetry for Lorentz transformations can also be appreciated in the following way. Suppose, $d > 0$. Then, as seen from $S$, any rigid body accelerating uniformly, undergoes a screw motion with the angular velocity about the screw axis directed along the direction of acceleration. Reversing the direction of time reverses each velocity but not acceleration. So now, after time reversal, angular velocity is directed opposite to acceleration, thus violating time reversal symmetry. Similarly for $d < 0$. The only way out is $d = 0$.

Since we have established that $d = 0$, from Eq. (17)-(20) it follows that $S''$ is parallel to $S$ and hence $S'$. So we can make $S'$ and $S''$ identical by choosing appropriate units in $S'$. To have compatibility with conventional notation, from now on we will use $S'$ instead of $S''$.

### e. 4th step: LT

The reciprocity principle (RP) states that the relative velocities of two frames with respect to each other are equal and opposite (Appendix). By virtue of PR, physics looks the same from all inertial frames. Observably therefore, there can be no absolute or innate identifier that distinguishes one frame from the other; in other words, no frame is preferred. Any transformation matrix taking one frame to the other thus should be a function of the velocity of the latter relative to the former only. By RP and PR, Eq. (17)−(20) remain valid if primed and unprimed coordinates are interchanged and $v$ is replaced by $-v$. Hence,

$$y' = ay \text{ and } y = ay' \qquad (22)$$

so that

$$a^2 = 1. \qquad (23)$$

Since no boost, i.e. $v = 0$, implies $y = y'$ we must have $a = 1$ instead of $-1$. To find the remaining scalar coefficients, note that

$$\begin{pmatrix} x' \\ t' \end{pmatrix} = \begin{pmatrix} -c & cv \\ fv & e \end{pmatrix} \begin{pmatrix} x \\ t \end{pmatrix} \text{ and } \begin{pmatrix} x \\ t \end{pmatrix} = \begin{pmatrix} -c & -cv \\ -fv & e \end{pmatrix} \begin{pmatrix} x' \\ t' \end{pmatrix}. \qquad (24)$$

Therefore,

$$\begin{pmatrix} -c & -cv \\ -fv & e \end{pmatrix} = \begin{pmatrix} -c & cv \\ fv & e \end{pmatrix}^{-1} = \begin{pmatrix} e & -cv \\ -fv & -c \end{pmatrix}/D, \qquad (25)$$

where

$$D = -ce - cfv^2. \qquad (26)$$

Comparing elements of the leftmost and rightmost matrix in Eq. (25), we have $-c = e$ and $D = 1$. So, Eq. (26) becomes

$$e^2 + efv^2 = 1. \qquad (27)$$

To complete the transformation equations we must now look for $e$ and $f$ as functions of $v^2$. Note that $e$ is dimensionless. $e(v^2)$ can be dimensionless non-trivially, only when there is an inherent speed scale $K$ (say) in every frame. When $f = 0$, we see from Eq. (27) that $e$ is trivially dimensionless as $e^2 = 1$. In that case $e = 1$, since $x = x'$ and $t = t'$ for $v = 0$ (i.e. no boost). This is the Galilean transformation (GT).

For nonzero $f$, however, note that any two events simultaneous in $S'$ or $S''$, have their difference in $x$ coordinates and time $t$ related by Eq. (20) as

$$0 = e\Delta t + fv\Delta x. \qquad (28)$$

Our discussion has no place for particles with infinite speed since they are more straight lines than particles. A particle is understood as having measure zero at any instant and a particle in one frame is also understood as a particle from any other. What more, a particle with finite speed cannot be finitely accelerated to have infinite speed and become a straight line. In other words, the concept of particle is such that particles can move with finite speeds only, i.e. a particle cannot be at two places simultaneously [7]. The two events simultaneous in $S'$ or $S''$ therefore, can never be part of the trajectory or world line of a particle. Hence, seen from $S$, no particle can traverse distance $\Delta x$ in time $\Delta t$ if they are related by Eq. (28). There must therefore, be a speed limit in $S$. By PR, such a speed limit must be universal, i.e. the same in all frames. This then is the supposed speed scale, $K$. Taking the unit of speed such that $K = 1$, all allowed speeds are less than or equal to one. From Eq. (28) then,

$$\left|\frac{e}{fv}\right| > 1. \qquad (29)$$

Time differentiating Eq. (17) using Eq. (20) we get the general velocity addition law which implies that a particle moving along $X$ or $X'$ axis with velocity $w$ is seen from the other frame to move with velocity

$$(w \mp v)/(1 \pm fvw/e). \qquad (30)$$

By virtue of Eq. (29) and $w, v < 1$, this expression (Eq. (30)) can be shown to be a monotonic increasing function of $w$. Hence, maximum speed in one frame is seen as maximum speed from the other; in other words,

$$(1 \mp v)/(1 \pm fv/e) = 1, \qquad (31)$$

or,

$$f = -e. \qquad (32)$$

From Eq. (27) now, we find that

$$e = 1/\sqrt{1 - v^2}. \qquad (33)$$

For completeness we rewrite the complete LT for a pure boost:

$$\mathbf{r}' = \mathbf{r} + \frac{(e-1)(\mathbf{v}\cdot\mathbf{r})\mathbf{v}}{v^2} - e\mathbf{v}t, \qquad (34)$$

$$t' = e(t - \mathbf{v}\cdot\mathbf{r}). \qquad (35)$$

### f. Some remarks

It is interesting to note the parallel between (or the complementarity of) Eq. (16) and (32). Both equations relate the coefficient of $x$ to that of $t$ in the transformation equations. To get Eq. (16) we considered events *co-local* in the primed frame, while to get Eq. (32) we required events *simultaneous* in the primed frame.

To get Eq. (32), some authors [8] refer to a third frame (that is in motion relative to both $S$ and $S'$ along their $X$ and $X'$ axes) and then assume that composition of two pure boosts along the same direction must be a pure boost, an assumption that is deemed redundant hereby. LT, as we saw, can indeed be derived without such a reference.

Most articles deriving LT without the light postulate [1] establish $e$ as

$$e = 1/\sqrt{1 - \frac{v^2}{\kappa}}, \qquad (36)$$

where $\kappa$ is some universal constant. According to Pauli [9] however, nothing can naturally be said about the sign, magnitude and physical meaning of $\kappa$. The usual critique is that it seems possible to derive the general form of the transformation formulae only but not their physical content. In this regard, our derivation of the transformations as presented here is not devoid of physical content or meaning as the concept of a universal speed limit naturally entered the discussion (Sec. (II.e)). The existence of such a limit was argued physically and then exploited to derive LT completely. GT with no inherent speed scale emerges here as qualitatively (physically) different from LT, while in conventional approaches with Eq. (36), GT is merely interpreted as a numerical choice ($\kappa = 0$) different from LT ($\kappa \neq 0$). Also note that our approach, unlike others, is also free from the onus of justifying the sign of the universal constant.

Also note that the setting up of parallel triads is absolutely inessential for the derivation of LT or GT. We could have totally eliminated the Sec (II.c) from our development. (We inserted it here merely to contrast our development with those by other authors). Whether the boosted frame is parallel or not to the unboosted frame is a consequence of the boost itself. It is preposterous to *construct* the boosted frame. The boost shall give the boosted frame as its logical conclusion, without any need for construction.

### III. CONCLUSION

In this paper we mainly focussed on discussing the need for time isotropy as a postulate to obtain the Lorentz transformation. Also, a parallel goal has been to provide a neat and somewhat novel derivation of Lorentz transformations. In such a derivation the concept of a finite speed limit enters naturally, and the role of it as a distinguishing factor between Lorentz and Galilean transformations becomes clarified. In the Appendix, a derivation of the reciprocity principle has also been presented. Care has been taken to keep the present article as self-contained as possible. However, for deeper discussions regarding the foundations of special relativity, the reader is referred to Ref. [10].

### APPENDIX: The Reciprocity Principle

The reciprocity principle has been proved from STH, SI and PR in Ref. [11]. We however, present here a somewhat different approach to it. Take any two inertial frames $S$ and $S'$ such that $S'$ is moving with respect to $S$ with velocity $\boldsymbol{v}$ and the transformation $S \to S'$ exists only by virtue of that motion. We will prove the reciprocity principle if we can show that relative velocity of $S$ with respect to $S'$,

$$\boldsymbol{v}' = -\boldsymbol{v} \tag{A1}$$

From spatial isotropy of $S$ and the fact that the transformation $S \to S'$ exits by virtue of the relative velocity of $S'$ only, we conclude that

$$\boldsymbol{v}' = \emptyset(v)\hat{\boldsymbol{v}}, \tag{A2}$$

where $v = |\boldsymbol{v}|$ and $\hat{\boldsymbol{v}}$ is the unit vector along $\boldsymbol{v}$. By principle of relativity now, it follows that

$$\boldsymbol{v} = \emptyset(v')\hat{\boldsymbol{v}}', \tag{A3}$$

where $v' = |\boldsymbol{v}'|$ and $\hat{\boldsymbol{v}}'$ is the unit vector along $\boldsymbol{v}'$. For $\emptyset(v) > 0$, $v' = \emptyset(v)$ and $\hat{\boldsymbol{v}}' = \hat{\boldsymbol{v}}$. Hence, from Eq. (A2) and (A3),

$$\boldsymbol{v} = \emptyset(\emptyset(v))\hat{\boldsymbol{v}}, \tag{A4}$$

i.e.,

$$\emptyset(\emptyset(v)) = v, \tag{A5}$$

If $\emptyset(v) < 0$, we can similarly show that

$$-\emptyset(-\emptyset(v)) = v, \tag{A6}$$

or, by substituting $\theta$ for $-\emptyset$,

$$\theta(\theta(v)) = v, \tag{A7}$$

Eq. (A5) and (A7) give the same solution for $\emptyset$ and $\theta$. Henceforth, therefore, we concentrate on Eq. (A5) only.

We assume that the domain of $\emptyset$ is an open interval on the non-negative real number line. Moreover, as seen from Eq. (A5), $\emptyset(\emptyset(v))$ is a regular function of $v$ and hence, differentiable for every value of $v$. Differentiating both sides of Eq. (A5), we get

$$\frac{d\emptyset(v)}{dv}\Big|_{\emptyset(v)} \frac{d\emptyset(v)}{dv}\Big|_{v} = 1. \tag{A8}$$

If $\emptyset$ was discontinuous at any $v = v_0$, $\frac{d\emptyset(v)}{dv}\Big|_{v_0}$ would diverge, thus implying $\frac{d\emptyset(v)}{dv}\Big|_{\emptyset(v_0)} = 0$ from Eq. (A8). But $\frac{d\emptyset(v)}{dv} = 0$ at any $v$ would violate the one-to-one character of the mapping $\emptyset(v)$. Hence, $\emptyset$ must be continuous.

Continuity and bijectivity of the mapping $\emptyset(v)$ imply that it is a strictly increasing or decreasing function of $v$ [12]. Since we must have $\emptyset(0) = 0$, the condition $\emptyset(v) > 0$ implies that $\emptyset$ is strictly increasing with $v$. As shown in Ref. [11], a strictly increasing and continuous $\emptyset$, satisfying Eq. (A5), can only have one form, namely

$$\emptyset(v) = v. \tag{A9}$$

Solution of Eq. (A7) similarly gives

$$\emptyset(v) = -\theta(v) = -v. \tag{A10}$$

It can be argued that the first choice for $\boldsymbol{v}'$ viz. Eq. (A9), should be discarded since it leads to inconsistencies [13]. Hence, we are left with Eq. (A10), which is equivalent to Eq. (A1), and thus the reciprocity relation stands proved.

Our approach to the reciprocity principle has a distinct advantage over that of Ref. [11]. We have proved the continuity of the mapping $\emptyset$ instead of assuming it [14].